\documentclass[reprint, aps, pra]{revtex4-1}

\usepackage{subfigure}
\usepackage{amsmath}
\usepackage{amssymb}
\usepackage{enumitem}
\usepackage{graphicx}
\usepackage{dcolumn}
\usepackage{bm}
\usepackage{xcolor}
\usepackage{soul}
\usepackage[colorlinks=true,allcolors=blue]{hyperref}

\graphicspath{{GAAfigures/}}

\newcommand{\ket}[1]{\left| #1 \right>}
\newcommand{\bk}[2]{\left< #1 \middle| #2 \right>}

\begin{document}

\title{Generalized adiabatic approximation to the quantum Rabi model}

\author{Zi-Min Li$ ^{1,} $}
\email{Zimin.Li@anu.edu.au}

\author{Murray T. Batchelor$ ^{2,3,} $}
\email{murray.batchelor@anu.edu.au}
\affiliation{
	$ ^1 $Department of Theoretical Physics, Research School of Physics, Australian National University, Canberra ACT, 2601, Australia\\
	$ ^2 $Mathematical Sciences Institute, Australian National University, Canberra ACT, 2601, Australia\\
	$ ^3 $Centre for Modern Physics, Chongqing University, Chongqing 400044, People's Republic of China
}

%

\date{\today}

\begin{abstract}
The quantum Rabi model (QRM) describes the interaction between a two-level system (qubit) and a quantum harmonic oscillator. 
In the limit where the qubit frequency is smaller than the harmonic frequency, the QRM can be well approximated by the adiabatic approximation (AA). 
The AA is widely used due to its simplicity and explicit physical interpretation. 
However, the level crossings in the spectrum of the QRM predicted by the AA are determined by the zeros of Laguerre polynomials, which deviate from the exact points. 
We propose a new approximation to the QRM that predicts the level crossings correctly. 
This is done by exploiting a surprising connection between isolated exact solutions to the QRM and the Laguerre polynomials in the AA. 
We thus refer to this approach as the generalized adiabatic approximation (GAA). 
By construction, the GAA always predicts the exact exceptional spectrum 
and approximates the regular spectrum remarkably well in a much larger parameter regime than the AA. 
This generalized approach offers a framework to deal with the family of Rabi-type light-matter interaction models in a simple but accurate manner. 
\end{abstract}

\maketitle

\section{Introduction}

The quantum Rabi model (QRM) is one of the simplest models in quantum physics \cite{Rabi_1936,Jaynes_1963}. 
It describes the interaction between a two-level system and a harmonic oscillator. 
Despite its simplicity, the QRM exhibits rich physics and has found applications in various contexts, 
from quantum optics \cite{Vedral_2005} to solid state physics \cite{Wagner1986}. 
For example, in quantum optics, the QRM describes the interaction between a two-level atom and a linearly polarized radiation mode. 
On the other hand, in solid state physics, an electron hopping between two sites and interacting with a vibrational mode can be depicted by the same Hamiltonian \cite{Reik_1982}. 
In the past decade, the realization of the QRM in circuit quantum electrodynamics (cQED) systems is a particularly 
exciting development \cite{Forn_D_az_2019,Frisk_Kockum_2019,Blais_2021}. 
This is partly because the cQED systems have become one of the most promising candidates for quantum computation. 
In these setups, the QRM describes a superconducting qubit coupled to a circuit resonator \cite{Niemczyk2010,Yoshihara_2016}.

The simple form of the QRM has undoubtedly concealed the complexity of its solvability. 
After more than eighty-years, the QRM has only been solved in the past decade \cite{Braak_2011,Chen2012,Zhong_2013,Maciejewski2014,Xie_2017}. 
Moreover, there is still no simple closed-form expression for the general energy eigenvalues of the QRM. 
In the analytic solution, the energy spectrum of the QRM is classified into regular and exceptional parts \cite{Braak_2011}. 
The regular spectrum, to which most eigenvalues belong, is obtained through the zeros of transcendental functions, 
whereas the exceptional solutions are isolated and given by the zeros of constraint polynomials with the system parameters as variables \cite{Kus1985}. 
These exceptional solutions correspond to the level crossings in the energy spectrum of the QRM.

Due to the lack of simple closed-form solutions, a number of approximations to the regular spectrum of the QRM are still actively pursued. 
Apart from the celebrated rotating-wave approximation (RWA) defining the Jaynes-Cummings model \cite{Jaynes_1963},  
which is at the heart of light-matter interaction in the strong coupling regime, 
the adiabatic approximation (AA) \cite{Irish_2005} has also been extensively adapted by theorists and experimentalists 
alike to describe the ultrastrong and deep-strong coupling regimes. 
The well-known generalized rotating-wave approximation (GRWA) has been developed by combining the RWA and the AA, 
providing a framework to derive analytic solutions to the QRM and its relatives \cite{Irish_2007,Zhang_2013}. 
Also, the AA has been adapted to classify the coupling regimes in light matter interaction processes \cite{Rossatto_2017}.
Experimentally, the AA has been exploited to interpret the spectroscopic results obtained in circuit QED experiments \cite{Yoshihara_2018}. 
Several other approximations to the QRM have been developed, each tailored to a particular range of parameters \cite{Boite2020}.

However, none of the existing approximations accurately capture the level crossing points in the energy spectrum. 
For example, the RWA breaks down in the ultrastrong coupling regime, before the appearance of the first crossing points. 
The AA predicts the level crossings through the zeros of Laguerre polynomials, which are only exact when the qubit frequency vanishes. 
To overcome this limitation, we propose a simple-form approximation based on the displaced oscillator picture and the exact solutions 
to the exceptional energy spectrum. 
We refer to this approach as the generalized adiabatic approximation (GAA). 
By construction, this GAA always predicts exact level crossings, and surprisingly, it also performs remarkably well on approximating 
the regular spectrum. 
This generalized approach offers a framework to treat light-matter interaction models with isolated exact solutions \cite{Kus_1986}.

This paper is organized as follows. 
In Sec.~\ref{SectionAA} we review the AA and collect the results for level crossing points of the QRM. 
In particular, we discuss the Laguerre polynomials in the AA and the constraint polynomials defining the exact exceptional solutions. 
In Sec.~\ref{SectionGAA} we observe the similarity and difference in these results and obtain an approximation which outperforms 
the AA and expands the validity into non-perturbative regimes. 
We demonstrate the advantages of the GAA by comparing the approximations with the numerical results. 
Some further discussion is given in Sec.~\ref{SectionDiscussion} with concluding remarks in Sec.~\ref{SectionConclusion}.

\section{Adiabatic approximation and exceptional solutions}\label{SectionAA}

\subsection{Model Hamiltonian}

In terms of the Pauli matrices $\sigma_x$ and $\sigma_z$ for a two-level system with level splitting $\Delta$ the 
Hamiltonian of the QRM reads ($\hbar=1$) 
\begin{equation}\label{HQRM}
	H = \dfrac{\Delta}{2}\sigma_z + \omega a^\dagger a + g \sigma_x \left(a^\dagger + a\right) . 
\end{equation}
The quantum harmonic  oscillator, or the single mode bosonic field, is described by the creation and annihilation operators $a^\dagger$ and $a$ with frequency $\omega$. 
The interaction between the two systems is via the coupling $g$.

The QRM conserves parity of the excitation number, which corresponds to a $ \mathbb{Z}_2 $ symmetry \cite{Braak_2011,Xie_2017}. 
Indeed, it can be easily shown that Hamiltonian (\ref{HQRM}) commutes with the parity operator 
$ P =\sigma_z e^{i \pi a^\dagger a} $, namely, $ \left[ H , P \right] =0  $.  
This implies that the QRM Hamiltonian matrix can be diagonalized into two blocks, each belonging to a parity subspace. 
Eigenvalues from different symmetry sectors are allowed to cross due to this $ \mathbb{Z}_2 $ symmetry.
More specifically, crossings exist within level pairs in the energy spectrum of the QRM.

\subsection{Displaced oscillator picture and adiabatic approximation}

\begin{figure}[b]
	\includegraphics[width=.85\linewidth]{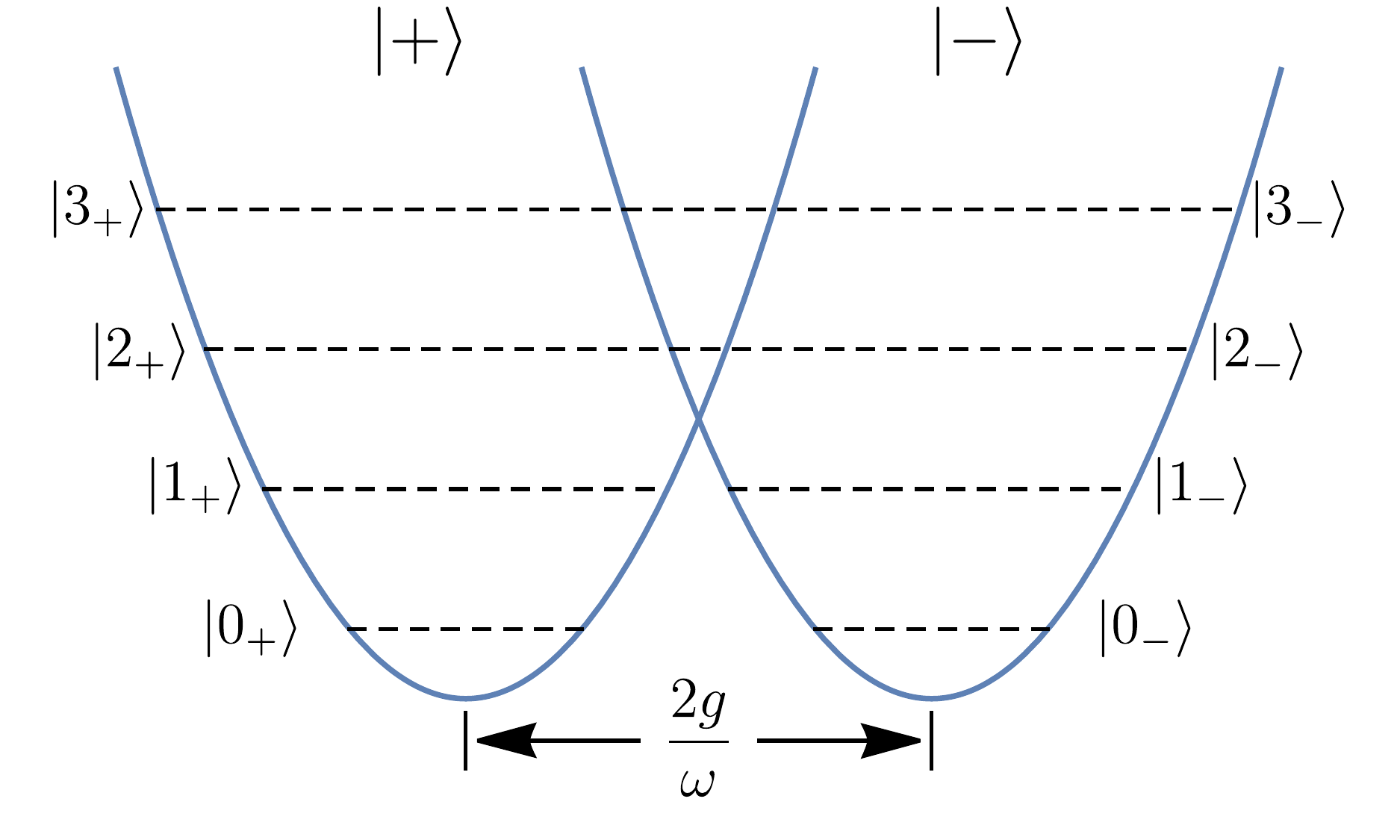}
	\caption{Sketch of the displaced oscillator picture for the QRM. When $\Delta=0$, the two oscillators are independent from each other, and the energy levels are doubly degenerate, as described by Eq.~(\ref{Edo}). When $ \Delta\ne 0 $, but the assumption $ \Delta/\omega\ll 1 $ holds, the qubit term $ \frac{\Delta}{2}\sigma_z $ induces tunneling between the degenerate eigenstates $ \ket{n_-,-} $ and $ \ket{n_+,+} $ and all other tunneling processes may be considered negligible. When $\Delta/\omega$ is large enough and the above assumption fails, higher order tunneling processes need to be taken into consideration. }
	\label{DisplacedOscillatorPicture}
\end{figure}

The QRM is trivially solvable in the atomic degeneracy limit $\Delta=0$, where the Hamiltonian reduces to
\begin{equation}\label{Hd0}
	H^{\Delta=0} = \omega a^\dagger a + g \sigma_x \left(a^\dagger + a\right) .
\end{equation}
In the basis of $ \ket{\pm x} $, where $\sigma_x \ket{\pm x} = \pm \ket{\pm x}$, 
we can rewrite the Hamiltonian in the form of displaced creation and annihilation operators, namely, 
\begin{equation}\label{Hd02}
	H^{\Delta=0} = \omega \left[ \left( a^\dagger \pm \dfrac{g}{\omega} \right) \left( a \pm \dfrac{g}{\omega} \right) \right] - \dfrac{g^2}{\omega}. 
\end{equation}
It follows that Hamiltonian (\ref{Hd0}) is diagonalized by a unitary transformation generated by the 
position displacement operator $\mathcal{D}(\alpha) = \exp[{-\alpha\left(a^\dagger - a\right)}]$, 
with displacement amplitude $\alpha = \pm g/\omega$ associated with the spin states. 
This gives rise to the solutions  
\begin{equation}\label{Edo}
	\begin{aligned}
			\Psi_{n,\pm}^\text{do} &= \ket{n_\pm,\pm }, \\	E_{n,\pm}^\text{do} &= n \, \omega - \dfrac{g^2}{\omega},
	\end{aligned}
\end{equation}
of Hamiltonian (\ref{Hd0}). Here $ \ket{n_\pm,\pm } = \ket{n_\pm}\otimes \ket{\pm x} $ with $ \ket{n_\pm} = \mathcal{D}(\pm g/\omega)\ket{n} $ 
being the displaced Fock states, also known as the generalized {or extended} coherent states \cite{Chen_2008,Philbin_2014}. 
Intuitively, Eq.~(\ref{Hd02}) can be understood as two displaced harmonic oscillators, 
with the displacement amplitude $ \pm g/\omega $ associated with the qubit states $ \ket{\pm x} $, 
as shown visually in Fig.~\ref{DisplacedOscillatorPicture}.
This interpretation is known as the displaced oscillator picture and has been extensively used in light-matter interaction models \cite{Irish_2014,Philbin_2014,Li2021,Li2021a}.

Next, we write the QRM Hamiltonian in the basis of displaced oscillator eigenstates (\ref{Edo}) and switch on the parameter $\Delta$. 
The qubit term $ \frac12{\Delta}\sigma_z $ couples these eigenstates. 
Equivalently, this coupling may be understood as the tunneling between the two displaced oscillators. 
The key assumption of the AA is that only the tunneling process between the eigenstates with the same $ n $ is considered, 
and higher order terms are neglected. 
By doing so, the QRM Hamiltonian is decomposed into block-diagonal form, consisting of infinitely many $ 2\times 2 $ blocks 
\begin{equation}\label{AAblock}
	H_n^\text{AA} =
	\begin{pmatrix}
		E_{n,+}^\text{do}  &  \frac{\Delta}{2} \bk{n_-}{n_+} \\
		\frac{\Delta}{2}\bk{n_-}{n_+}  &  E_{n,-}^\text{do}
	\end{pmatrix},
\end{equation}
which can be readily solved. 
The term $ \bk{m_-}{n_+} $ is the overlap between the generalized coherent states $ \ket{m_-} $ and $ \ket{n_+} $, given by $ (m\le n) $
\begin{equation}
	\bk{m_-}{n_+} = \exp\left[-{2\alpha^2}\right] |2\alpha|^{n-m}\sqrt{\dfrac{m!}{n!}}L_m^{n-m}\left(4\alpha^2\right).
\end{equation}
Here $ L_m^{n-m}(x) $ are the associated Laguerre polynomials. 
The eigenstates and eigenvalues of Eq.~(\ref{AAblock}) are
\begin{equation}\label{EAA}
	\begin{aligned}
		{\Psi_{n,\pm}^\text{AA}} &= \dfrac{1}{\sqrt{2}}\left(\ket{n_+,+} \pm \ket{n_-,-}\right), \\
		E_{n,\pm}^\text{AA} &= n\omega -\dfrac{g^2}{\omega} \pm \Omega_n^\text{AA}, 
	\end{aligned}
\end{equation}
where the tunneling strength is described by
\begin{equation}\label{AAOmega}
	\Omega_n^\text{AA} = \dfrac{\Delta}{2} \exp\left[-\dfrac{2g^2}{\omega^2}\right]L_n\left(\dfrac{4g^2}{\omega^2}\right),
\end{equation}
{where $ L_n(x)=L_n^0(x) $ are the standard Laguerre polynomials. }
The solutions in Eq.~(\ref{EAA}) are in a form as simple as that of the celebrated RWA. 
In this paper, we aim to maintain this simplicity with a higher degree of approximation.

The AA has been derived several times in different contexts in the literature, see, e.g., Refs.~\cite{Graham1984,Feranchuk_1996,Irish_2005}. 
Because it is in simple form and physically intuitive, the AA has been widely used both theoretically and experimentally. 
A generalized rotating-wave approximation (GRWA) was derived by combining the AA and the RWA \cite{Irish_2007}, 
leading to a framework to analytically deal with the QRM and its related models \cite{Zhang_2013}. 
In other work, the AA is adapted to define the so-called perturbative deep-strong coupling (DSC) regime \cite{Rossatto_2017}. 
On the experimental side, the AA has been used to explain ``the inversion of qubit energy levels'' in cQED setups operating in the DSC regime \cite{Yoshihara_2018}.

It is intuitive to interpret the AA in the displaced oscillator picture shown in Fig.~\ref{DisplacedOscillatorPicture}. 
When $\Delta=0$, the two harmonic oscillators are displaced by a distance of $\pm g/\omega$, respectively. 
If $\Delta$ becomes nonzero, it induces tunneling between the two oscillators. 
Under the condition $\Delta\ll \omega$, the level gaps between different oscillator levels are so large that 
the qubit energy quanta can never excite the oscillator states. 
Thus, the interaction or tunneling may be thought of as only occuring between the levels that have the same energy index $ n $. 
The tunneling strengths are then indicated by the overlap between the degenerate eigenstates of the oscillators, namely, the term $\Omega^\text{AA}_n$.

We can see from the form of Laguerre polynomials that the tunneling strengths are non-monotonic with respect to the displacement 
amplitude $ g/\omega $ \cite{Irish_2005}. 
Indeed, for certain values of $ g/\omega $, the tunneling strengths on some levels vanish, 
corresponding to the level crossings in the energy spectrum as a function of $ g $. 
To be specific, from Eq. (\ref{EAA}), the energy gap between the $ n $th pair of levels is given by
\begin{equation}\label{deltaE}
	\delta E_n^\text{AA} = E_{n,+}^\text{AA} - E_{n,-}^\text{AA} = \Delta \exp\left[-\dfrac{2g^2}{\omega^2}\right] L_n\left(\dfrac{4g^2}{\omega^2}\right). 
\end{equation}
Thus, the locations of the level crossing points, where the energy gap is zero, 
are determined by the zeros of Laguerre polynomials $ L_n(4g^2/\omega^2) $. 
Since $\Delta$ does not enter the arguments of the Laguerre polynomials, 
it is obvious that the level crossings predicted by the AA are independent of $\Delta$.

We note that the tunneling strengths between non-degenerate levels are never strictly zero unless $\Delta=0$. 
In this regard, the AA is only exact in the limit $\Delta/\omega\rightarrow 0^+$. 
If $\Delta$ becomes large and comparable with $\omega$, the performance of AA worsens. 
A manifestation of this is that the level crossings predicted by the AA substantially deviate from the exact ones. 
In the displaced oscillator picture, this means that the tunneling between non-degenerate levels is no longer negligible. 
In this case, the tunneling cannot be simply described by the Laguerre polynomials $ L_n $. 
Effectively, the two potentials are no longer harmonic. 
Furthermore, if $ \Delta>\omega $, the AA breaks down by introducing unphysical level crossings into the energy spectrum in the regime $ g/\omega <0.5 $. 
We further illustrate these arguments in Sec.~\ref{SectionGAA}.

After the above analysis, the amendment to the AA naturally relies on taking into account the effect of $\Delta$. 
We now turn to obtain some intuition from the exact solutions for the exceptional spectrum of the QRM.

\begin{table*}[t]
	\centering
	\begin{tabular}{p{.05\textwidth}<{\centering} p{.3\textwidth}<{\centering} p{.6\textwidth}<{\centering}}
		\toprule
		&& \\[-.7em]
		$ n $ & Laguerre polynomials $ L_n(4g^2) $ & Normalized constraint polynomials $ K_n (g ,\Delta) $\\
		&& \\[-.7em]
		\colrule
		&& \\[-.7em]
		0& 1 & 1\\
		&& \\[-.7em]
		1& $1 -  4g^2 $ & $ 1- 4g^2 - \dfrac{\Delta^2}{4} $\\
		&& \\[-.7em]
		2& $ 1 -8 g^2 + 8 g^4 $& $ 1 -8 g^2 + 8 g^4 -\dfrac{5 \Delta ^2}{16}+\dfrac{3 \Delta ^2 g^2}{4} + \dfrac{\Delta ^4}{64} $ \\
		&& \\[-.7em]
		3& 			$ 1-12g^2 + 24g^4 -\dfrac{32g^6}{3} $ & 
		$ 1-12g^2 + 24g^4 -\dfrac{32g^6}{3} -\dfrac{49 \Delta ^2}{144} +\dfrac{29g^2\Delta^2}{18} -\dfrac{11g^4\Delta^2}{9} + \dfrac{7\Delta^4}{288} -\dfrac{g^2\Delta^4}{24} - \dfrac{\Delta^6}{2304}$\\
		&& \\[-.7em]
		\botrule
	\end{tabular}
	\caption{The first few Laguerre polynomials and normalized constraint polynomials defined in Eq.~(\ref{Kn}). }
	\label{tablepoly}
\end{table*}

\subsection{Exact solutions for the exceptional spectrum}

As pointed out in the Introduction, there are no simple closed-form solutions to the regular spectrum of the QRM. 
Instead, only the level crossing points in the spectrum can be determined through finite-term constraint polynomials of system parameters \cite{Li_2016,Kimoto_2020}. 
For this reason, the QRM is also referred to as quasi-solvable \cite{Zhang2013,Batchelor_2015,Guan_2018}. 
These isolated solutions are called Juddian points, and the determining polynomials have been derived via different approaches \cite{Judd_1979,Reik_1982,Kus1985,Moroz_2018}. 
The simplest way to derive the constraint polynomials is probably through Bargmann representation of bosonic operators, as described in Ref.~\cite{Kus_1986}. 
In the following analysis, we set $\omega=1$ for simplicity. 

The constraint polynomials, denoted by $ P_n^n(g,\Delta) $, obey the  recurrence relation
\begin{equation}\label{JuddRecursion}
	\begin{aligned}
		&P^n_0(g,\Delta) = 1,\quad P^n_1(g,\Delta) = 4g^2 + \dfrac{\Delta^2}{4} - 1, \\
		&P^n_k(g,\Delta) = \left(4 k g^2 + \dfrac{\Delta^2}{4} - k^2 \right) P^n_{k-1}(g,\Delta) \\
		& \phantom{P^n_k =}- 4k(k-1)(n-k+1)g^2 P^n_{k-2}(g,\Delta).
	\end{aligned}
\end{equation} 
The zeroes of $P^n_n(g,\Delta)$ determine the degenerate points of the $n$th pair of levels in the QRM. 
Note that $ n=0 $ corresponds to the ground state and the first excited state. 

Here we now consider the pair of levels with $ n=2 $ as an example to demonstrate how the constraint polynomials 
are used to solve for the  level crossing points. 
From the recurrence relation (\ref{JuddRecursion}), we have 
\begin{equation}\label{P22}
	P_2^2(g,\Delta) = 4 - 32g^2 +32g^4 - \dfrac{5\Delta^2}{4} + 3g^2 \Delta^2 +\dfrac{\Delta^4}{16}.
\end{equation}
If we further fix $\Delta = 1.2 $, the equation $ P_2^2(g,\Delta) = 0 $ then gives two positive solutions for $g$, 
which are approximately 0.3074 and 0.8778, which can be confirmed by the crossing points in the energy spectrum displayed in Fig.~\ref{GAAlowB}. 

From the constraint polynomials, the number of degenerate points on each level is also known. 
To be specific, a mathematically proven theorem \cite{Kus1985,Li_2015} 
states that there are $ n-k $ level crossing points on the $ n $th level for $\Delta$ in the range 
\begin{equation}\label{DeltaTheorem}
	2k<\Delta/\omega < 2(k+1),
\end{equation}
with positive integer $ k $. 
In other words, there are more level crossings in higher levels. 
This fact is also partly true for Laguerre polynomials in the AA.
However, as already stressed, the dependence on $\Delta$ of the number of crossings cannot be correctly treated there. 

It is interesting to note that level crossing points are determined by certain polynomials both in the AA and the exact solutions. 
Our next task is to find the possible connection between Laguerre polynomials and the constraint polynomials.

\section{Generalized adiabatic approximation}\label{SectionGAA}

\subsection{Laguerre polynomials versus constraint polynomials}

We know from the constraint polynomials in Eq.~(\ref{JuddRecursion}) that the positions of the level crossing points 
in the spectrum depend on the value of $\Delta$, in contrast to the predictions of the AA. 
Interestingly, if we set $\Delta=0$ and eliminate the terms containing $\Delta$ from the constraint polynomials, 
we arrive at the Laguerre polynomials up to an integer overall coefficient. 

We again take the case of $ n=2 $ as an example. 
The corresponding Laguerre polynomial is 
\begin{equation}
	L_2(4g^2) = 1- 8g^2 + 8g^4. 
\end{equation}
Meanwhile, with $\Delta=0$, the constraint polynomial Eq.~(\ref{P22}) becomes 
\begin{equation}
	P_2^2(g,0) = 4-32g^2+32g^4 = 4 L_2(4g^2).
\end{equation}
In this case, the Laguerre polynomial $ L_2(4g^2) $ and the constraint polynomial $ P_2^2(g,0) $ share the same zeros, 
and thus give the same level crossing points. 
This comparison justifies the fact that the AA is only exact in the limit $\Delta\rightarrow 0^+$. 
In fact, the larger $\Delta$ is, the larger the deviation of the AA. 

The $\Delta$ terms in the constraint polynomials, e.g., in Eq.~(\ref{P22}), 
may be understood as the correction terms accounting for the tunneling processes between non-degenerate levels in the displaced oscillator picture. 
These correction terms become dominant when $\Delta$ is comparative with or even larger than $\omega$, where the assumption of the AA is invalid. 

Since both the regular and exceptional spectrum of the QRM are described by Laguerre polynomials in the AA, 
we expect that the constraint polynomials, which include the correction terms, can approximate the regular eigenvalues with a better performance. 
To this end, we need to properly correct the AA tunneling strength $\Omega_n^\text{AA}$ in Eq.~(\ref{AAOmega}).


\begin{figure*}[t]
	\subfigure{	
		\includegraphics[width=.317\linewidth]{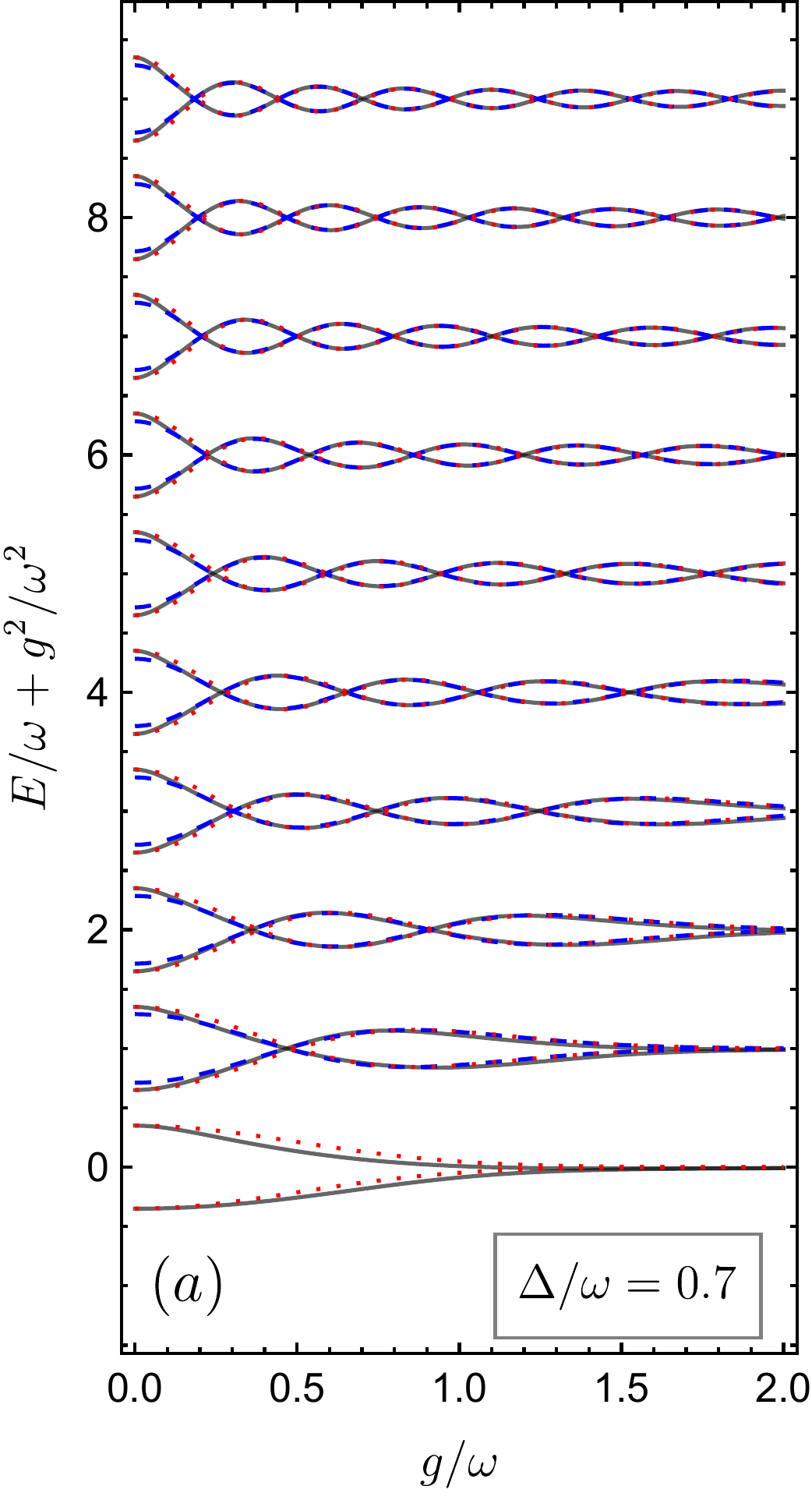}
		\label{GAAlowA}
	}
	\subfigure{	
		\includegraphics[width=.317\linewidth]{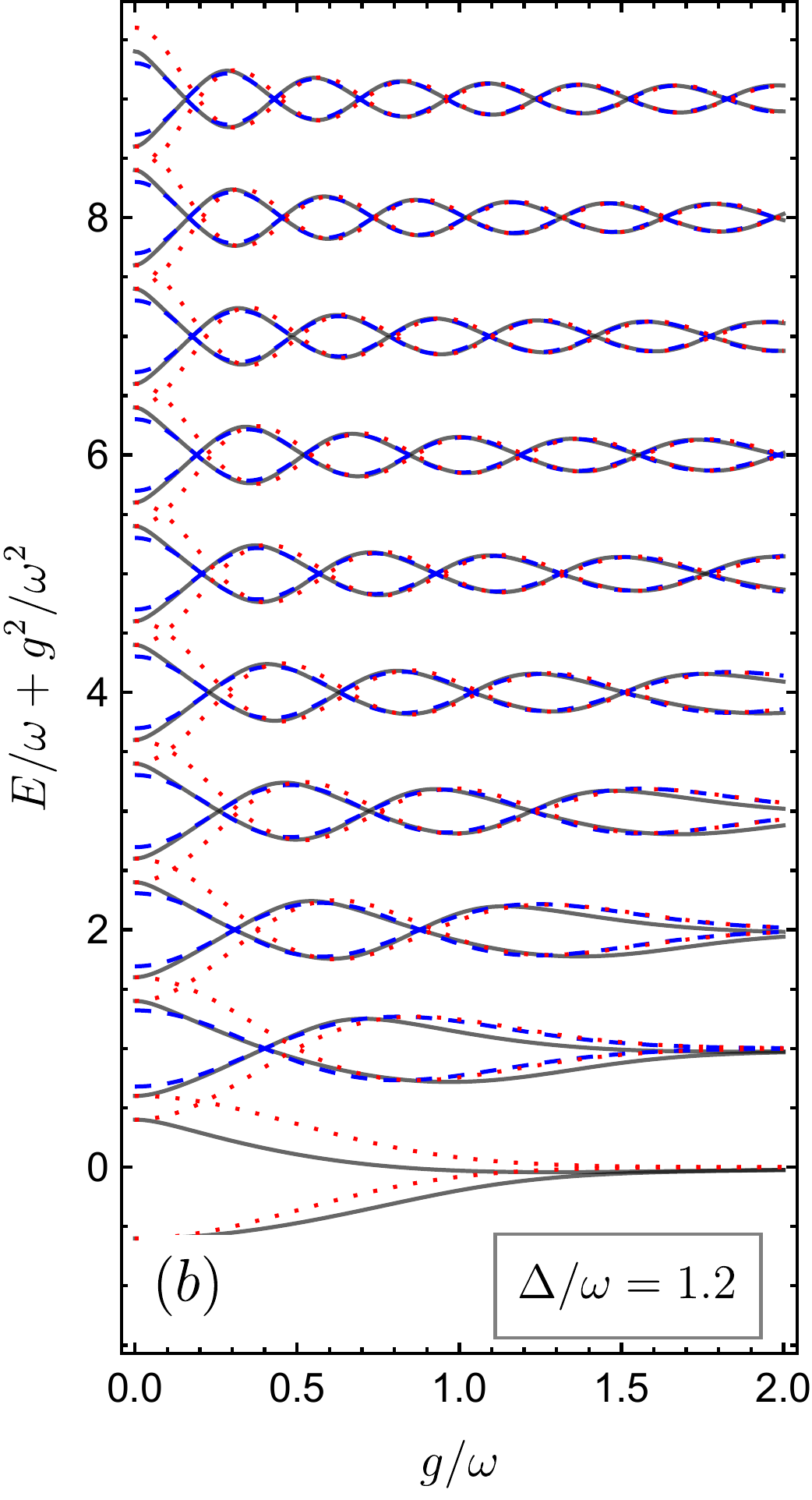}
		\label{GAAlowB}
	}
	\subfigure{	
		\includegraphics[width=.317\linewidth]{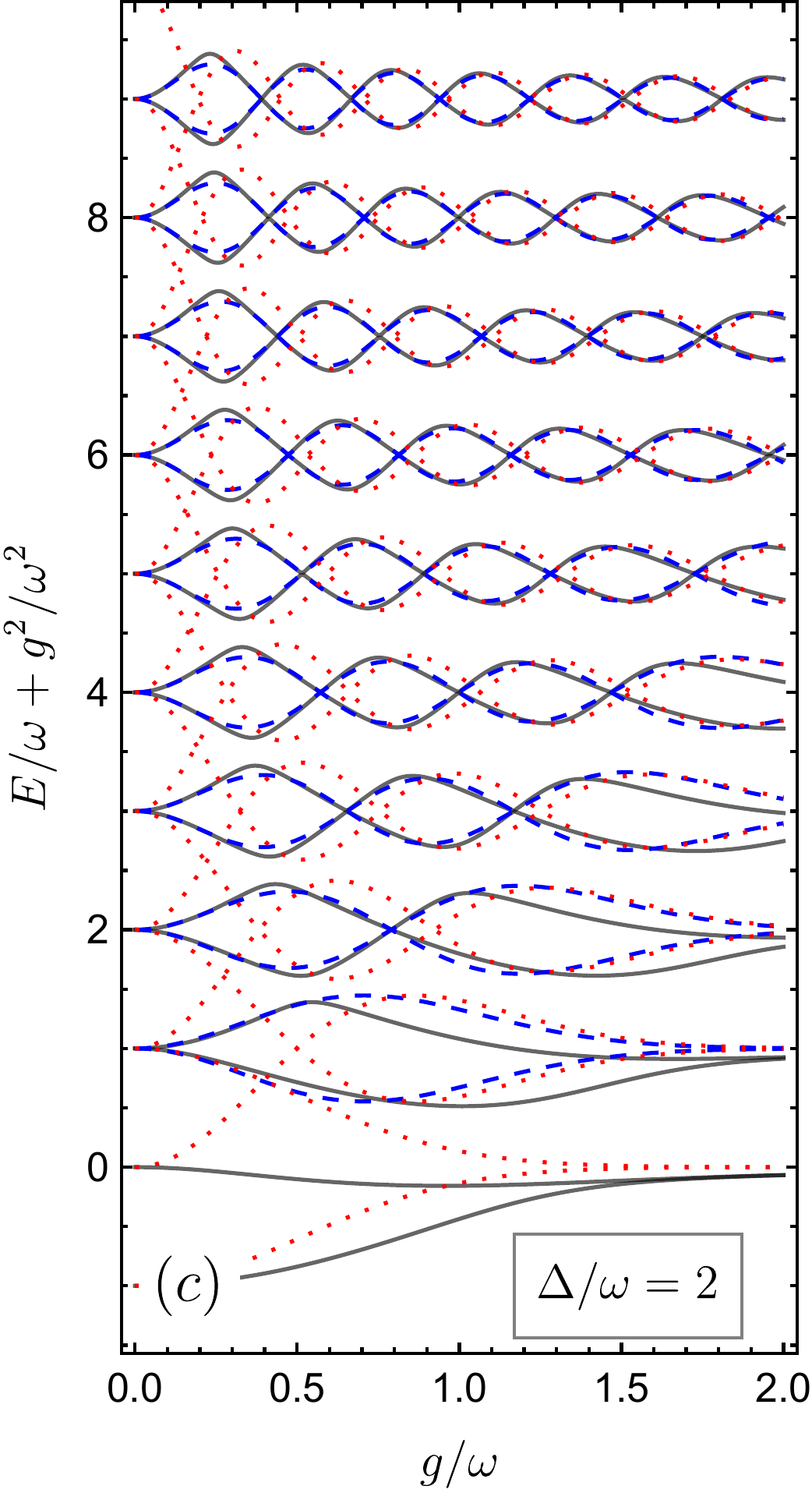}
		\label{GAAlowC}
	}
	\\
	\subfigure{	
		\includegraphics[width=.4\linewidth]{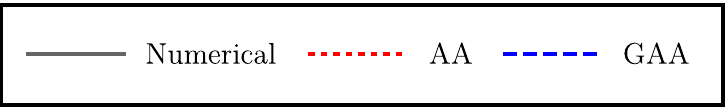}
	} 
	\caption{(a)-(c) Low energy levels of the QRM with respect to coupling $g/\omega$ obtained through exact diagonalization (grey solid), AA (red dotted) and GAA (blue dashed), with (a) $\Delta/\omega=0.7$, (b) $\Delta/\omega=1.2$ and (c) $\Delta/\omega=2$.  {In (a) both the AA and GAA work very well within the perturbative regime. In (b) the AA fails by introducing unphysical crossings before the first real crossing points. In (c) the AA completely fails whereas the GAA works remarkably well.} }
	\label{GAAlow}
\end{figure*}

\subsection{Corrected tunneling strengths}

With $ P_n^n(g,0)/L_n(4g^2) $ being an integer, it is natural to propose the normalized constraint polynomials
\begin{equation}\label{Kn}
	K_n(g,\Delta) = \dfrac{P^n_n(g,\Delta)}{P^n_n(0,0)}.
\end{equation}
The first few Laguerre polynomials $L_n(4g^2)$ and normalized constraint polynomials $K_n(g,\Delta)$ are listed in Table~\ref{tablepoly}.
It is easy to confirm $K_n(g,0) = L_n(4g^2)$, which justifies the choice of the normalization factor in Eq.~(\ref{Kn}).

Until this point, substantial improvements can already be obtained by simply replacing the Laguerre polynomials in Eq.~(\ref{EAA}) 
with the normalized constraint polynomials in Eq.~(\ref{Kn}). 
By doing so, the exceptional eigenvalues are exact and the regular spectrum is also seen to be approximated with very good agreement.

Further corrections can be obtained by going back to the displaced oscillator picture. 
The two oscillators are displaced by a distance of $ \pm g/\omega $, which becomes exact when $\Delta=0$.
However, if $\Delta$ is nonzero, the tunneling leads to an effective displacement on the oscillators \cite{Li2021}. 
This effect due to $\Delta$ is neglected in the AA. 
Meanwhile, the constraint polynomials can be regarded as the Laguerre polynomials plus the effect of extra displacements due to $\Delta$. 
Therefore, we are in a position to seek more accurate displacements $\alpha_n(g,\Delta)$ such that $L_n(4\alpha_n^2) = K_n(g,\Delta)$, 
predicting the exact level crossings. 
Note that we use the subscript in $\alpha_n$ to account for the fact that the oscillators are no longer harmonic. 
It turns out to be highly non-trivial, or even impossible, to find such exact displacement amplitudes $\alpha_n$. 
The fundamental reason is that the displaced oscillator picture is never exact for $\Delta$ nonzero. 
Thus, we cannot expect exact results from an approximating treatment. 
However, it is still meaningful to look for displacement amplitudes $\alpha_n$ which are more accurate than $g/\omega$ 
to improve the AA and the subsequent results.

Our strategy is to find corrected arguments for the Laguerre polynomials such that $L_n(4\alpha_n^2)$ have the exact $ \Delta^{2n} $ terms. 
Careful observation of Table~\ref{tablepoly} indicates that the corrected arguments $\alpha_n$ should then satisfy
\begin{equation}\label{newarg}
	4\alpha_n^2 =
	\begin{cases}
			 4g^2 + \dfrac{\Delta^2}{4\sqrt[n]{n!}},& n \in \mathbb{N^+}, \\
			 4g^2, & n=0.
	\end{cases}
\end{equation}
{This correction can be easily derived through the recurrence relation as shown in Appendix A. }
It can be confirmed that the other $\Delta$ terms in the new Laguerre polynomials $L_n(4\alpha_n^2)$ are also very close to those in $K_n(g,\Delta)$. 
Therefore, $\alpha_n$ can be regarded as a good approximation to the real displacement amplitudes. 
We note that when $n=0$, there is no $\Delta$ term in $L_0(4g^2)$ and thus no correction in this case.

The corrected arguments (\ref{newarg}) immediately lead to the corrected tunneling strengths
\begin{equation}\label{GAAOmega}
	\Omega_n^\text{GAA} = \dfrac{\Delta}{2}\exp\left[-2\alpha_n^2\right]K_n(g,\Delta),
\end{equation}
which is the key result of the GAA.
Note that in Eq.~(\ref{GAAOmega}) we stick to exact polynomials $K_n(g,\Delta)$ rather than the ``nearly exact'' $L_n(4\alpha_n^2)$. 
{A comparison between these two options is discussed in Sec.~{\ref{Comparison2Options}}. }

\begin{figure*}[!t]
	\subfigure{	
		\includegraphics[width=.317\linewidth]{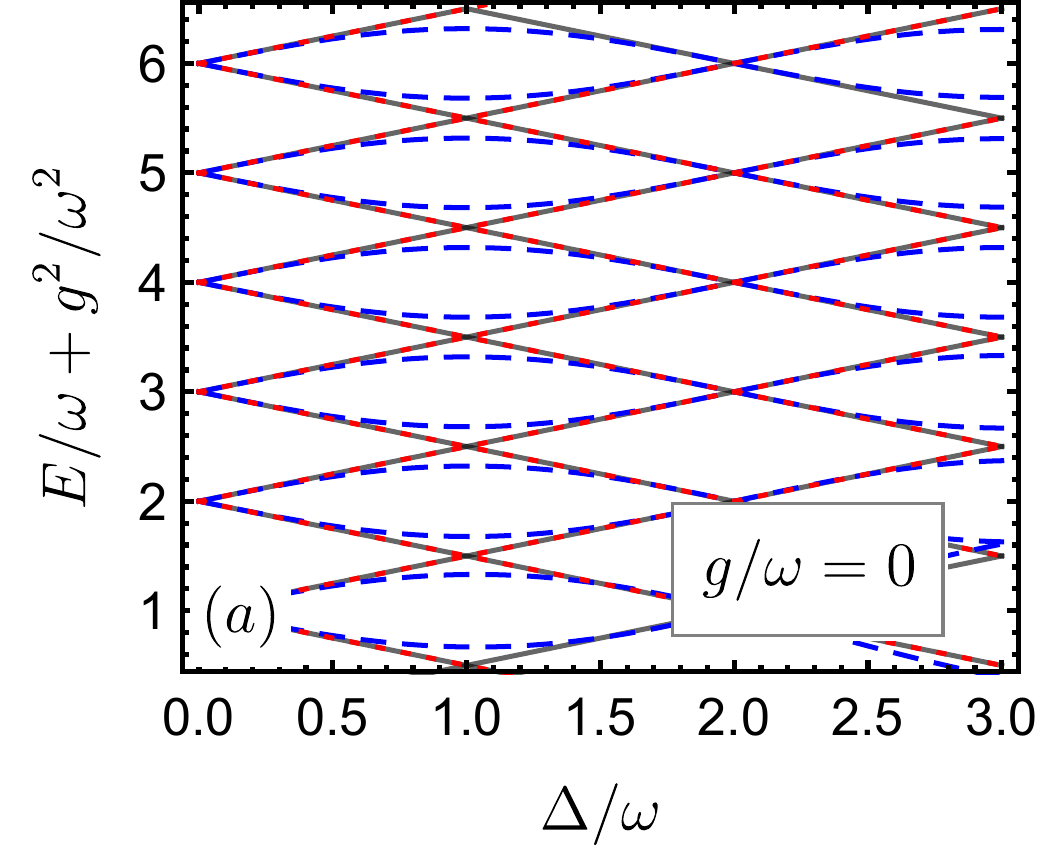}
		\label{Edeltalow}
	}
	\subfigure{	
		\includegraphics[width=.317\linewidth]{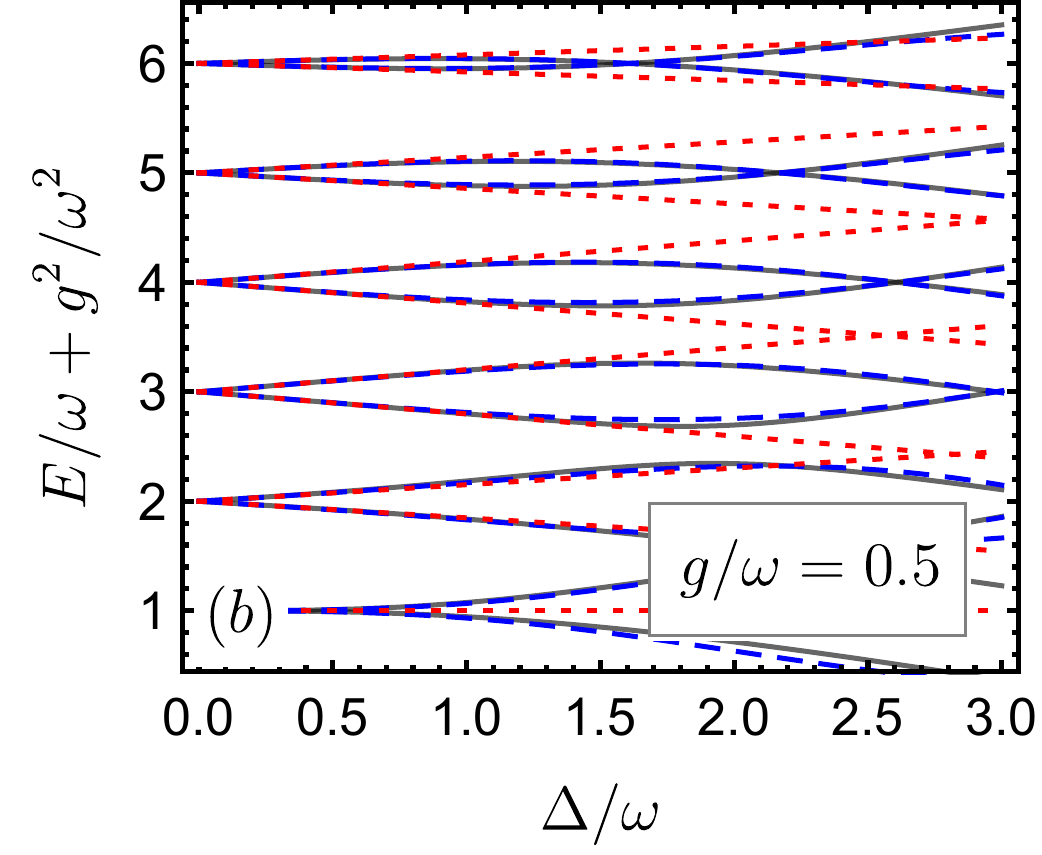}
		\label{Edeltalow2}
	}
	\subfigure{	
		\includegraphics[width=.317\linewidth]{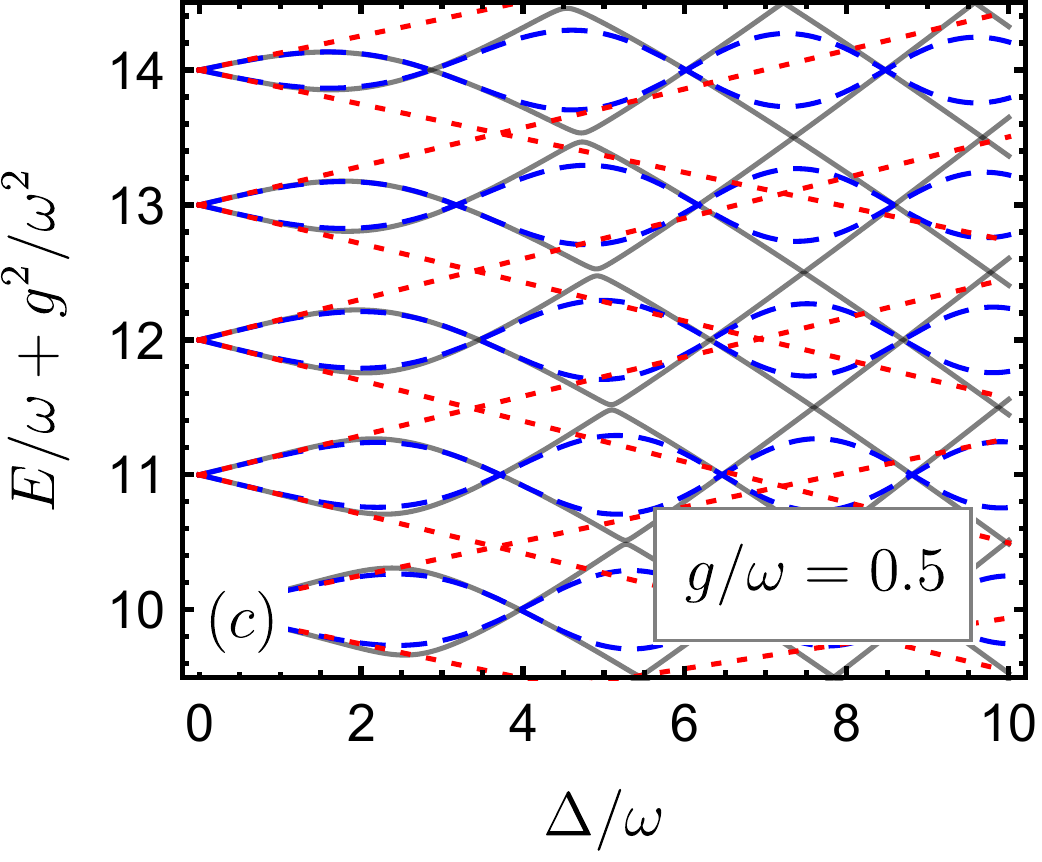}
		\label{Edeltahigh}
	}
	\\
	\subfigure{	
		\includegraphics[width=.4\linewidth]{GAAlegend.pdf}
	} 
	\caption{Energy levels of the QRM with respect to $\Delta/\omega$ obtained through exact diagonalization (grey solid), AA (red dotted) and GAA (blue dashed). 
		(a) The decoupled limit $ g/\omega=0 $. The AA is exact in this limit while the GAA deviates around odd integer values of $\Delta$. 
		(b) Low levels with $ g/\omega=0.5 $. The AA fails for large $\Delta/\omega$ whereas the 
		GAA captures the overall dependence on $\Delta$ very well. 
		(c) Higher levels with $ g/\omega=0.5 $. The GAA captures the overall dependence on $\Delta$ but fails to describe the strong avoided crossings for extreme values of $\Delta/\omega$. 
	} 
\end{figure*}

\subsection{Energy spectrum}

It is now straightforward to write down the eigenvalue expressions under the GAA. 
We replace the tunneling strengths $\Omega_n^\text{AA}$ in Eq.~(\ref{AAOmega}) with the corrected ones $\Omega_n^\text{GAA}$ in Eq.~(\ref{GAAOmega}) to obtain the eigenvalues
\begin{equation}\label{EGAA}
		E_{n,\pm}^\text{GAA} = n \, \omega -\dfrac{g^2}{\omega} \pm \Omega_n^\text{GAA}(g,\Delta). 
\end{equation}
In this way, the energy eigenvalues are corrected without destroying the simplicity of the AA. 

We turn now to calculating the energy spectrum and check the improvements offered by the GAA. 
The lowest energy levels of the QRM with different values of $\Delta$ are shown in Figs.~\ref{GAAlowA}-\ref{GAAlowC}. 
For comparison, we also show the spectrum obtained through exact diagonalization.
We already know that the AA works reasonably provided $\Delta/\omega<1$ but breaks down beyond this regime. 
Indeed, from Fig.~\ref{GAAlowA} with $\Delta/\omega=0.7$, we can see that both AA and GAA approximate the exact eigenvalues with good agreement. 
Only marginal improvement can be found in the GAA around level crossing points. 
When $\Delta/\omega =1.2$, as displayed in Fig.~\ref{GAAlowB}, 
the adiabatic approximation breaks down by introducing unphysical level crossings that do not exist in the exact energy spectrum. 
Moreover, the existing level crossings are also obviously out of phase. 
In Fig.~\ref{GAAlowC} where $\Delta/\omega = 2$, the improvements are even more profound. 
The AA completely breaks down in all finite coupling regimes while the GAA approximates the overall spectrum with a remarkable agreement. 
We note that there are no level crossings between the ground state and the first excited state, thus the GAA brings no improvement to the lowest two energy levels. 
This exception can also be deduced from the fact that $P_0^0 = L_0(4g^2) =1$, which does not depend on $\Delta$. 
Thus, for the sake of clarity, the lowest two levels are not shown in the figures. 
In summary, substantial improvements to the energy levels offered by GAA can be seen.

Since the positions of degenerate points are always exact in the GAA, the level crossings may be regarded as calibration points. 
It is therefore natural to expect that the GAA works even better for higher levels of the QRM due to the increasing number of level crossings. 
{This can be seen in} the higher energy levels of the QRM in {Fig.~{\ref{GAAlow}}}. 
For $\Delta/\omega=1.2$, the AA induces unphysical crossings and deviates from the exact degenerate points, whereas the GAA works very well, as anticipated. 
In {Fig.~{\ref{GAAlow}}, for} $ \Delta/\omega=2 $, the assumption of the AA completely breaks down and the energy levels wildly cross each other. 
The GAA, however, still works remarkably well and all the characteristic properties of the energy levels of the QRM remain.

Having seen that the GAA outperforms the AA in the large $\Delta$ regime, 
we are in a position to explore the validity of the GAA with respect to the parameter $\Delta$. 
We first consider the decoupled limit $ g=0 $, which is another trivial case of the QRM, with 
\begin{equation}
	H^{g=0} =  \dfrac{\Delta}{2}\sigma_z  + \omega a^\dagger a.
\end{equation}
The eigenvalues for the uncoupled case are simply $ E^{g=0}_{n,\pm} = n \, \omega \pm \Delta/2 $. 
The energy spectrum in the decoupled limit is displayed in Fig.~\ref{Edeltalow}.  
The AA predicts the correct eigenvalues in this limit. 
However, this correctness leads to unphysical crossings when $\Delta/\omega>1$, as can be observed in Figs.~\ref{GAAlowB}. 
The GAA, on the other hand, is not always exact in this limit, but the overall approximating behavior is rather good. 
When $ g=0 $, the constraint polynomials reduce to the nice form
\begin{equation}\label{Pnng0}
	P_n^n(0,\Delta) = \prod_{k=1}^n\left(\dfrac{\Delta^2}{4} - k^2\right), 
\end{equation}
correctly predicting the degeneracies when $ \Delta/\omega $ takes even integer values. 
It is non-trivial to obtain a similar closed-form expression for the general case $ P_n^n(g,\Delta) $.

The lowest energy levels with fixed parameter $ g/\omega=0.5 $ are displayed in Fig.~\ref{Edeltalow2}, 
from which we observe that the GAA performs remarkably well in a large parameter regime. 
The AA, on the other hand, fits well for small $\Delta$ but deviates substantially for large $\Delta$. 
From the AA results in Eq.~(\ref{EAA}), eigenvalues are linearly dependent on $ \Delta $ if other parameters are fixed, 
which is not true in the exact results. 
We also expect that the large number of level crossings in higher levels help guarantee the performance of the GAA. 
The higher levels in an even larger parameter regime are displayed in Fig.~\ref{Edeltahigh}. 
In the displaced oscillator picture, extremely large $\Delta$ means very strong and complicated tunneling processes between the two oscillators. 
Indeed, the AA completely breaks down for $\Delta/\omega>2$ in most coupling regimes. 
Owing to the rich crossing points, the overall eigenvalue structure is still captured by the GAA. 
In Fig.~\ref{Edeltahigh}, apart from the real crossings predicted by GAA, 
we also observe many strong avoided crossings in the regime where $\Delta/\omega>5$. 
These avoided crossings cannot be depicted by the current approximation and higher order corrections are needed.

\section{Further discussion}\label{SectionDiscussion}

\subsection{Validity}

Although the GAA is seen to work very well in a large parameter regime, it is still meaningful to discuss the validity of the approximation. 
We consider the cases where the GAA does not provide any improvements. 

The GAA always offers a remarkable description of the energy levels containing crossing points due to the corrected tunneling strength 
$ \Omega_n^\text{GAA} $ in Eq.~(\ref{GAAOmega}).
However, there are no level crossings in the lowest pair of levels, i.e., the ground state and the first excited state. 
In this case, $ K_0(g,\Delta) = L_0(4g^2) = 1 $, implying no correction from $\Delta$. 
Therefore, the GAA brings no improvement to the ground state and the first excited state. 
For completeness, we note that these states can be well described by the non-orthogonal qubit states in all parameter regimes \cite{Leroux_2017,Li2021}. 

More generally, the description of the lowest $ k $ pairs of energy levels cannot be well improved by the GAA for $\Delta$ in the range $2k<\Delta/\omega<2(k+1)$. 
This is due to the result given in Eq.~(\ref{DeltaTheorem}). 
The number of crossing points on level $ n $ is dependent on the value of $\Delta$. 
Therefore, for extremely large $\Delta$, the lowest levels are separated and no crossings exist.

{Regarding another aspect of the validity, we note that the GAA may also predict unphysical level crossings in some extreme parameter regimes, 
e.g., when $ \Delta/\omega>5 $ or $ g/\omega>2 $. This is due to the same reason as that in the AA. Specifically, the Hilbert space of the GAA is 
split into infinitely many two-dimensional invariant subspaces, and eigenvalues from different sectors are allowed to cross. 
This problem can be solved to some extent by looking for a more precise set of $ \alpha_n $ in {Eq.~(\ref{newarg})}. 
However, in the parameter regimes of interest, the GAA is simple and accurate enough. }

\subsection{Further discussion on $ L_n(4\alpha_n^2) $}\label{Comparison2Options}

\begin{figure}[t]
	\subfigure{
		\includegraphics[width=\linewidth]{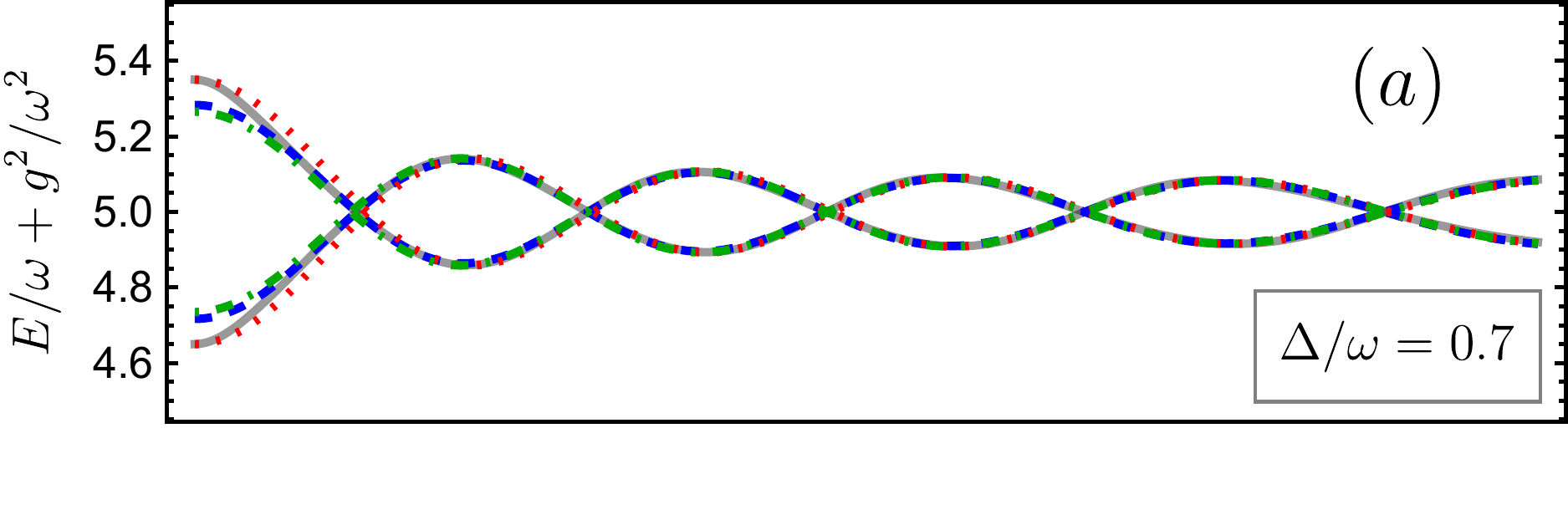}
		\label{LK1}
	}
	\\[-2.35em]
	\subfigure{
		\includegraphics[width=\linewidth]{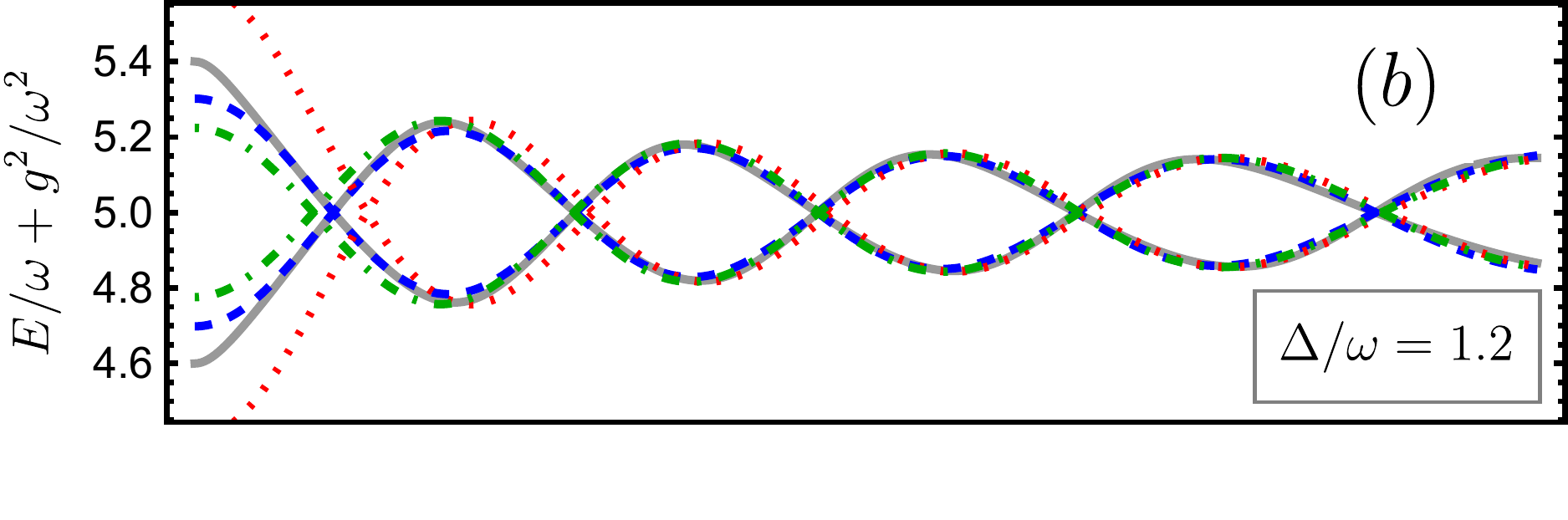}
		\label{LK2}
	}	
	\\[-2.35em]
	\subfigure{
		\includegraphics[width=\linewidth]{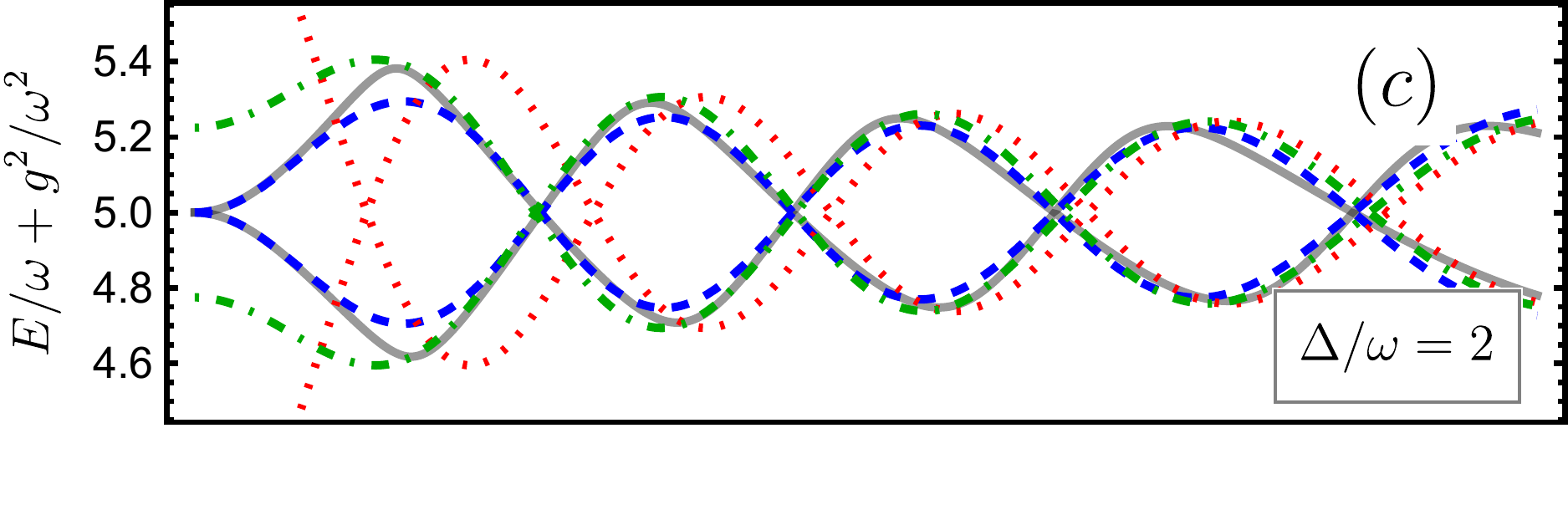}
		\label{LK3}
	}	
	\\[-2.35em]
	\subfigure{
		\includegraphics[width=\linewidth]{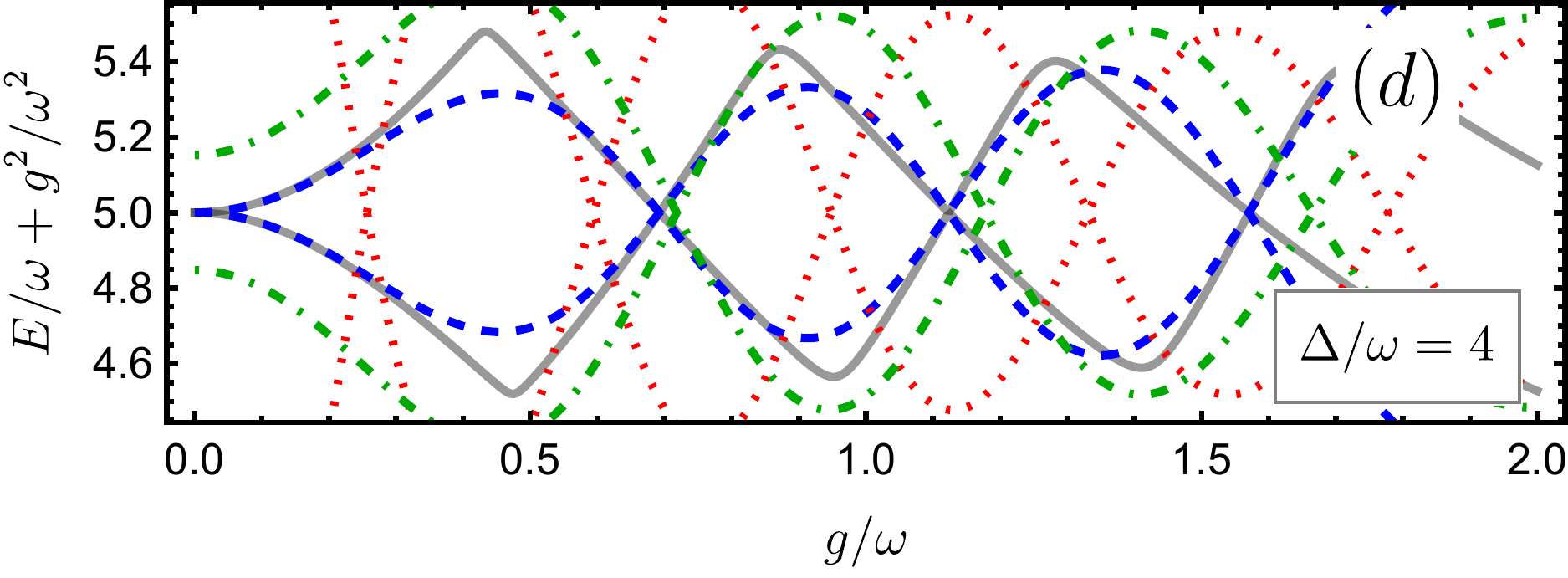}
		\label{LK4}
	}
	\subfigure{
		\includegraphics[width=.6\linewidth]{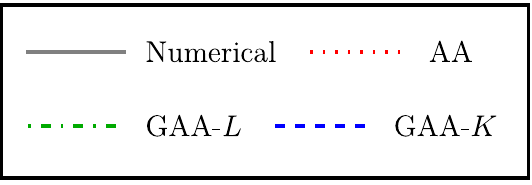}
	}
	\caption{A comparison between the two options, $ L_n(4\alpha_n^2) $ and $ K_n(g,\Delta) $, in the expression of the GAA. The exact numerical results and the AA are also included as benchmarks. }
	\label{LnKn}
\end{figure}

{In Eq.~{(\ref{newarg})} we obtained the $ n $-dependent displacement amplitudes $\alpha_n(g,\Delta)$. 
However, in the corrected tunneling strengths ~{(\ref{GAAOmega})}, we choose the exact normalized constraint polynomials $ K_n(g,\Delta) $ 
rather than the ``nearly exact'' Laguerre polynomials $ L_n(4\alpha_n^2) $. 
Here we briefly discuss the difference between these two options, which we refer to as GAA-$K$ and GAA-$L$ for simplicity. 

We take the level pair with $ n=5 $ as an example and calculate the eigenvalues through the two different options. 
The numerically exact results and the AA are also included for comparison. 
The results are displayed in Fig.~{\ref{LnKn}}. 
The case $\Delta/\omega=0.7$ is demonstrated in Fig.~{\ref{LK1}}, where both the AA and the GAA are valid. 
In this case, there is no noticeable difference between the two options of the GAA. 
As shown in Fig.~{\ref{LK2}} and Fig.~{\ref{LK3}}, when $\Delta/\omega=1.2$ and $ \Delta/\omega $, 
the AA deviates largely from the exact numerical results while both the two GAA options give reasonably good overall performance. 
In these cases, however, we notice that GAA-$L$ has some minor deviations around the level crossings. 
On the other hand, GAA-$K$ always has exact level crossings by construction. 
The deviation of GAA-$L$ can be even larger if we increase the value of $\Delta/\omega$, as shown in Fig.~{\ref{LK4}}.

We conclude this comparison by pointing out that GAA-$K$ gives the best performance in calculating the eigenvalues. 
However, GAA-$L$ comes with a form quite similar to the AA and thus may be more suitable to use in some unitary transformations. 
}

\subsection{Potential applications to other models}

The exact exceptional solutions exist in many light-matter interaction models \cite{Kus_1986}. 
Thus, the present GAA offers a framework to approximate these systems with both simplicity and accuracy. 
An important example is the biased or asymmetric quantum Rabi model (AQRM), 
where the $ \mathbb{Z}_2 $ symmetry is broken by a bias term \cite{Braak_2011}. 
In the AQRM, the level crossings are generally lifted apart from some special cases.
However, the isolated exceptional solutions exist even without level crossings in the spectrum or any obvious symmetry in the Hamiltonian. 
The corresponding constraint polynomials and their properties have been explored in detail \cite{Zhong_2014,Li_2015,Kimoto_2020}. 
We expect the present GAA can approximate the AQRM and, importantly, recover the conical intersections in the spectrum \cite{Batchelor2015}.

The anisotropic Rabi model, an interpolation between the Jaynes-Cummings model and the QRM, 
is another system that may benefit from the results presented in this paper. 
The AA has been applied to the anisotropic QRM \cite{Tomka_2014} and its isolated exceptional solutions have also been discussed in detail \cite{Chen_2021}. 
Particularly, for some parameters, there are level crossing points between the lowest two levels, in contrast to the present QRM. 
A combination of these results should in principle give rise to a better approximation to the model, including the ground state and the first excited state.

Since the GAA outperforms the AA in a very large parameter regime, it is also reasonable to expect improvements to the results derived using the AA.
The celebrated RWA breaks down in the ultrastrong coupling regime ($ g/\omega>0.1 $), 
and various new approximations have been proposed to fix this problem. 
Among them is the generalized rotating-wave approximation (GRWA) which works well for arbitrarily strong coupling strength near resonance. 
The GRWA is derived by first writing the QRM in the basis of AA eigenstates, and then applying the RWA \cite{Irish_2007}. 
The GRWA works well for arbitrary coupling strength $ g $, but it fails for $\Delta/\omega>1$ for the same reason that the AA fails. 
It is to be expected that the GRWA based on the GAA will lead to a better performance in a larger regime of $\Delta$. 
{Comparison between the GRWA and the GAA is shown in Appendix B.}

\section{Conclusion}\label{SectionConclusion}

In this paper, we propose the GAA to treat the QRM in a simple but accurate manner. 
The basic idea is to combine the perturbative AA and the isolated exact solutions to the exceptional spectrum of the QRM. 
This is achieved through the key observation that the constraint polynomials of system parameters reduce to 
Laguerre polynomials of the AA in the unperturbed limit $\Delta=0$. 
Thus, the constraint polynomials are considered as Laguerre polynomials corrected by the higher 
order tunneling processes in the displaced oscillator picture. 
By construction, the GAA not only predicts the level crossing points exactly, 
but also approximates the general spectrum with remarkably good agreement in large parameter regimes. 
Importantly, our results expand the perturbative treatment into non-perturbative regimes. 

The GAA offers a theoretical framework to analytically treat a family of light-matter interaction models where isolated exact solutions exist. 
The asymmetric and anisotropic QRM are discussed as examples for further work.
This approximation scheme is especially useful where the level crossings play an important role, 
as, e.g., near the conical intersection points of the AQRM.

\begin{acknowledgments}
	It is a pleasure to thank David Tilbrook and Devid Ferri for helpful discussions.
	This work is supported by the Australian Research Council through Discovery Grants DP170104934 and DP180101040. 
\end{acknowledgments}

\appendix

\section{Derivation of the corrected displacement}

{From the recurrence relation {(\ref{JuddRecursion})}, we know that the highest order $\Delta$ term is $ \left( {\Delta}/{2}\right)^{2n} $. 
On the other hand, from the compact expression {(\ref{Pnng0})} in the limit $ g=0 $, we have}
\begin{equation}
	P_n^n(0,0) = \left(-1\right)^n\prod_{k=1}^{n}k^2 = \left(-1\right)^n\left(n!\right)^2. 
\end{equation}
{It follows that the highest order $\Delta$ term in the normalized constraint polynomial is}
\begin{equation}\label{h1Delta}
	h_1(\Delta)=\left(-1\right)^n\dfrac{\Delta^{2n}}{4^n\left(n!\right)^2}. 
\end{equation}

{We are looking for a variable of the Laguerre polynomials of the form}
\begin{equation}
	x(g,\Delta) = 4\alpha_n^2(g,\Delta) = 4g^2 + f_n(\Delta),
\end{equation}
{where $ f_n(\Delta) $ is a polynomial in $\Delta$ such that the resulting Laguerre polynomials share the same highest order $\Delta$ term as the normalized constraint polynomials. 
The highest order $\Delta$ term can be determined through the closed form of the Laguerre polynomials, given by}
\begin{equation}
	L_n(x) = \sum_{k=0}^{n}\begin{pmatrix} n\\k \end{pmatrix}\dfrac{\left(-1\right)^k}{k!}x^k.
\end{equation}
{With $ g=0 $, the highest order $\Delta$ term is}
\begin{equation}\label{h2Delta}
	h_2(\Delta) = \dfrac{(-1)^n}{n!}f_n(\Delta)^n.
\end{equation} 
{On comparing $ h_1(\Delta) $ and $ h_2(\Delta) $, we have}
\begin{equation}
	\dfrac{(-1)^n}{n!}f_n(\Delta)^n = \left(-1\right)^n\dfrac{\Delta^{2n}}{4^n\left(n!\right)^2},
\end{equation}
{giving rise to the result}
\begin{equation}
	f_n(\Delta) = \dfrac{\Delta^2}{4\sqrt[n]{n!}}.
\end{equation}
{Finally, we obtain the corrected displacement amplitudes $\alpha_n$ given in Eq.~{(\ref{newarg})}.} 

\section{Comparison with the generalized rotating wave approximation}

\begin{figure*}[t]
\subfigure{\includegraphics[width=0.3\linewidth]{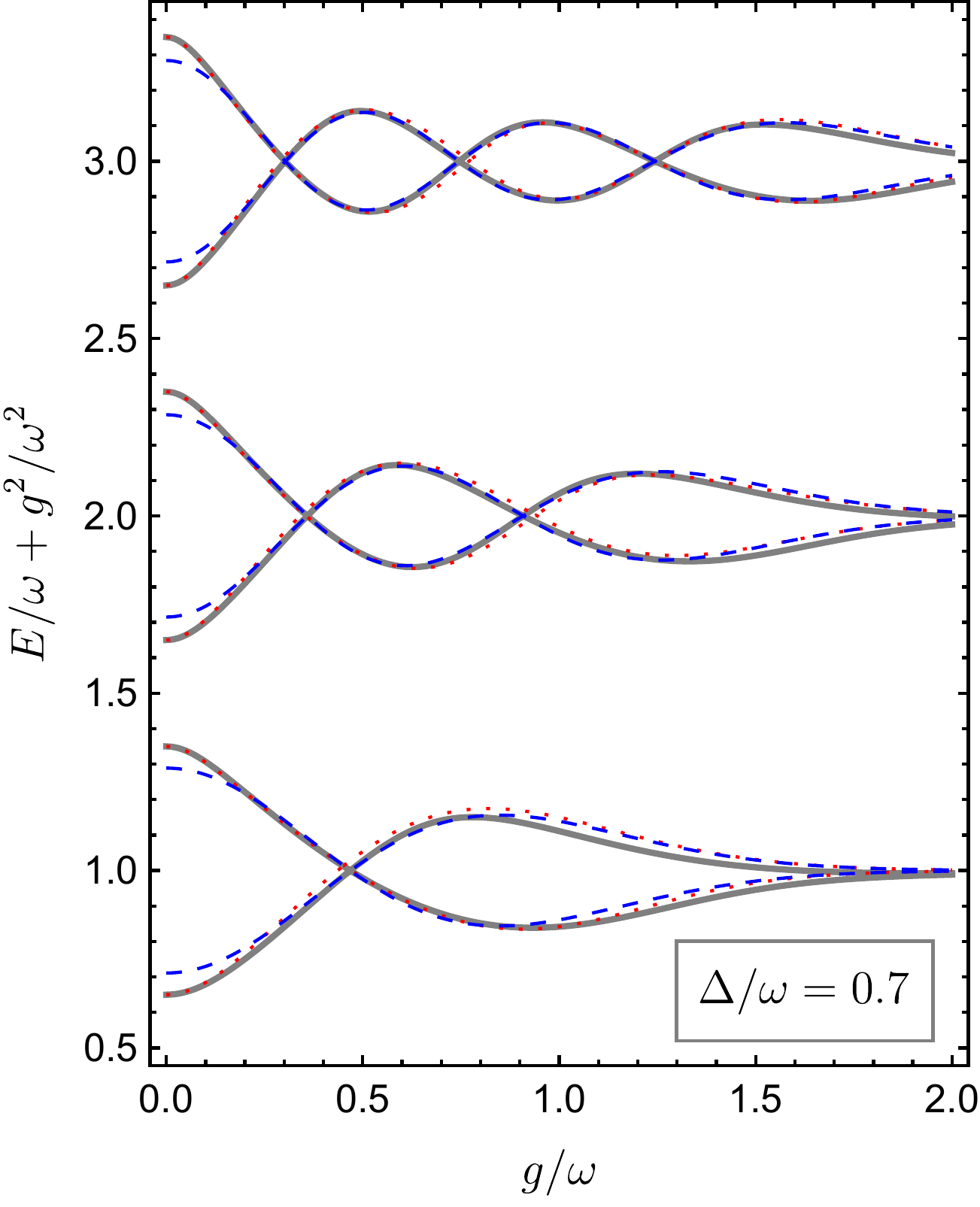}}
\subfigure{\includegraphics[width=0.3\linewidth]{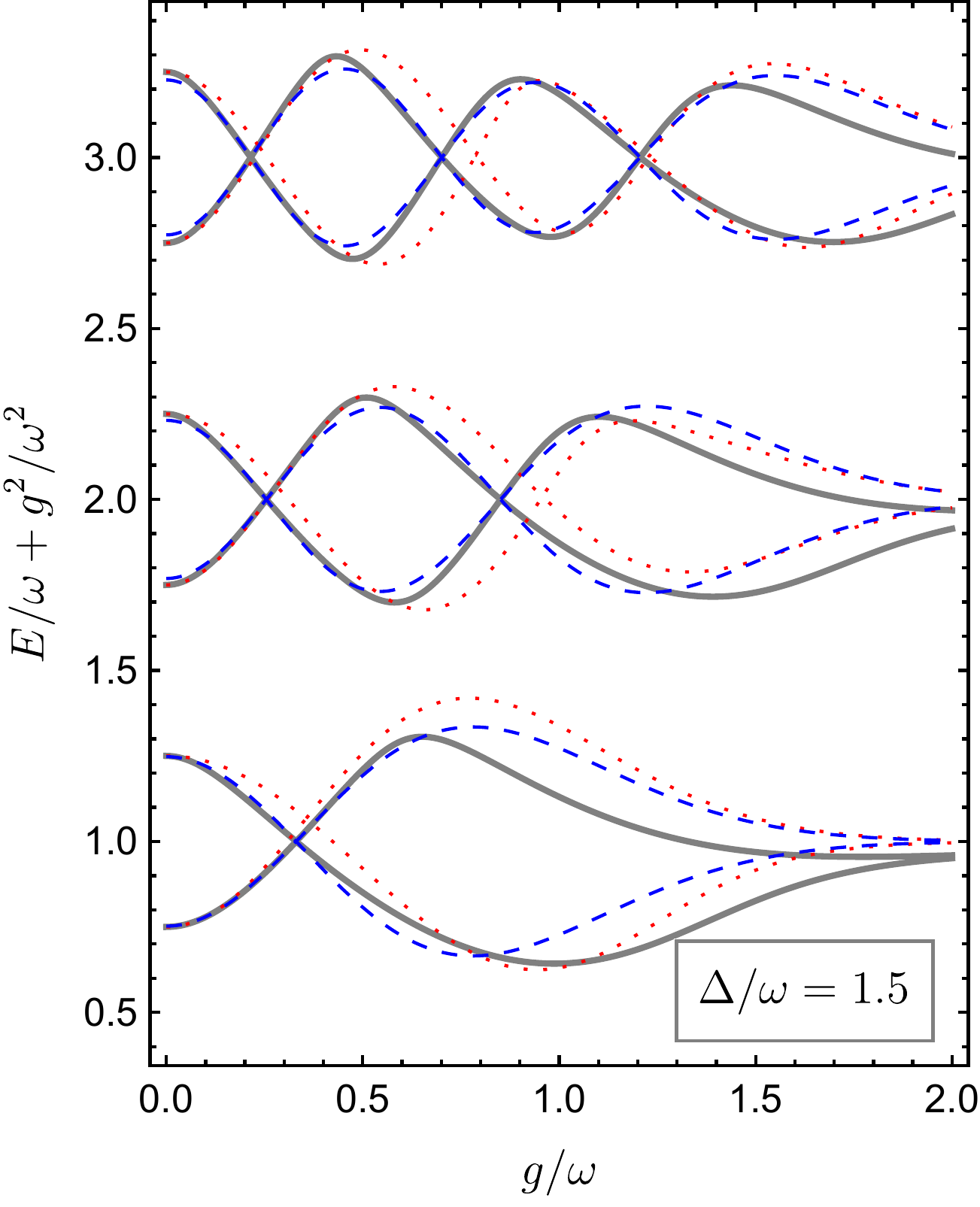}}
\subfigure{\includegraphics[width=0.3\linewidth]{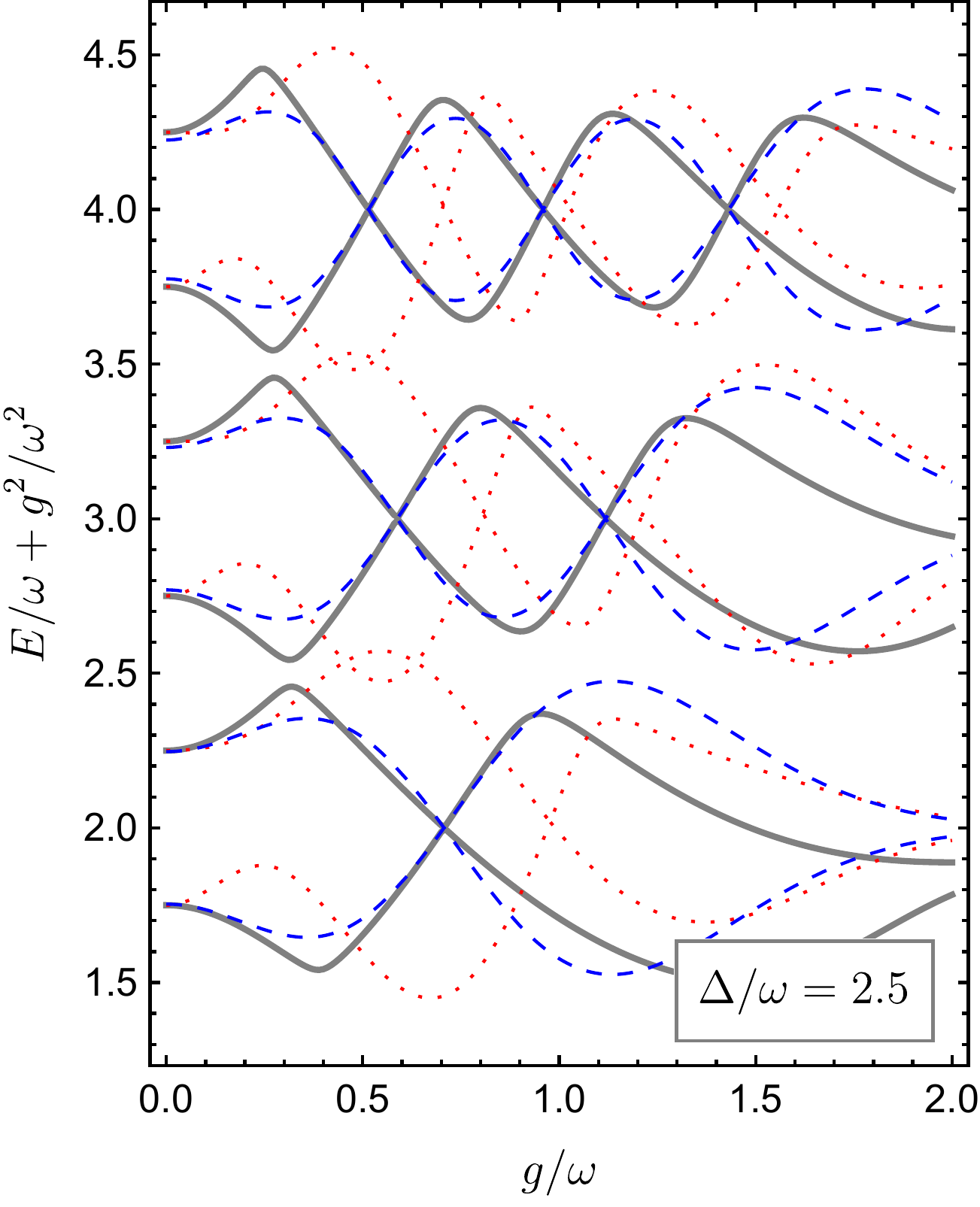}}
\subfigure{\includegraphics[width=0.4\linewidth]{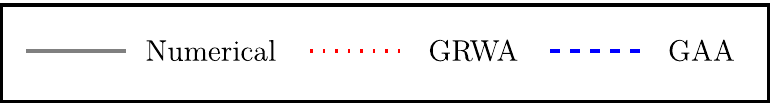}}
\caption{Comparison between the GRWA and the GAA with different values of $\Delta/\omega$. }
\label{GRWAvsGAA}
\end{figure*}

{The GRWA {\cite{Irish_2007}} describes the QRM with arbitrary coupling strength $ g $, provided that $\Delta/\omega$ is near unity.
Beyond this limit of $ \Delta $, the GRWA breaks down.
Moreover, unphysical crossings are expected in the GRWA since it also decomposes the Hilbert space into two-dimensional subspaces.

In Fig.~{\ref{GRWAvsGAA}}, we compare the GRWA with the GAA to justify the advantage of our method in a much larger parameter regime.}


%

\end{document}